\documentclass[11pt]{aastex}
\usepackage{emulateapj5}
\usepackage{onecolfloat}

\slugcomment{Version: 24 September, 2001, accepted by ApJL}

\shorttitle{Nulling Observations of Herbig Ae Stars}
\shortauthors{Hinz, Hoffmann, and Hora}

\begin{document}
\def\head{

\title{Constraints on Disk Sizes Around Young Intermediate-Mass Stars: \\  Nulling Interferometric Observations of Herbig Ae Objects\footnotemark[1]}

\footnotetext[1]{Observations reported here were obtained at the MMT Observatory, a joint facility of the University of Arizona and the Smithsonian Institution }


\author{Philip M. Hinz$^2$, William F. Hoffmann$^2$, and  Joseph L. Hora$^3$}

\affil{$^2$Steward Observatory, University of Arizona, 933 N. Cherry Ave., Tucson, AZ 85721, phinz@as.arizona.edu, whoffmann@as.arizona.edu}

\affil{$^3$ Harvard-Smithsonian Center for Astrophysics, 60 Garden Street, MS 42, Cambridge, MA 02138, jhora@cfa.harvard.edu}

\begin{abstract} 
Physical models of circumstellar disks surrounding pre-main sequence stars are currently constrained primarily by the spectral energy distribution (SED) of the system. Here we present first results from a survey of nearby Herbig Ae stars using nulling interferometry. We measure the spatial extent of the mid-IR emission for the stars HD 150193, HD 163296, and HD 179218.  The results indicate the size of mid-infrared emission around Herbig Ae stars may be much smaller than current models predict.  The observations constrain $\sim$90\% of the 10 $\micron$ flux to be within an approximately 20 AU diameter region, providing significant spatial limits for disk models. These results demonstrate the scientific potential for nulling interferometry to study circumstellar disks with better spatial discernment than is possible with standard mid-infrared imaging.
\end{abstract}

\keywords{stars:pre-main sequence --- stars:individual(HD150193, HD163296, HD 179218) --- techniques:interferometric}
}
\twocolumn[\head]

\section{Introduction}

for emulateapj
\footnotetext[1]{Observations reported here were obtained at the MMT 
Observatory, a joint facility of the University of Arizona and the Smithsonian Institution }

Circumstellar dust disks, the presumed nursery for planetary systems, are most easily studied while they are still young and are best seen around the brightest pre-main sequence stars. Herbig Ae/Be stars are known to harbor an excess of infrared (IR) emission above that of their photospheres, and recent evidence has pointed to this excess being in the form of a circumstellar disk (Natta et al. 2001). However, significant differences have been highlighted between the Ae and Be spectral types indicating that while Herbig Ae stars do indeed harbor disks, as evidenced by detection of their millimeter flux, Herbig Be stars (earlier than B8), in general, do not(Natta et al. 2000). Herbig Ae stars are also interesting as the evolutionary precursors of the Vega-excess stars first discovered by the IRAS satellite to have infrared excess due to a debris disk.

The spectral energy distribution (SED) has been a powerful tool for probing these circumstellar disks. Simple, geometrically thin, optically thick models of a disk fail to reproduce the observed spectral energy distribution, underestimating the amount of flux in the mid and far-IR.  Recent models (Kenyon and Hartmann 1987; Chiang and Goldreich 1997; D'Alessio et al. 1998) that can reproduce the SED predict a significant flare to the radial structure of the disk due to the vertical hydrostatic equilibrium, increasing the disk's exposure to the starlight.  Absorption of this stellar flux heats the surface layer of the flared disk to a higher temperature than the interior.  In addition to creating emission features, the mid-IR flux from such systems is expected to be significantly more extended, compared to geometrically thin disk models.

In this paper we present spatial constraints on disks around three Herbig Ae stars using a new technique called nulling interferometry. 

\section{Nulling Interferometry}
Nulling interferometry is a novel technique for detecting faint circumstellar structure around bright point sources. The technique overlaps the pupils of two telescopes or subapertures of a single telescope to interferometrically suppress the light from an on-axis star. Light from an angular distance of
$\theta=\lambda/(2\,b)$ away will constructively interfere, allowing detection of faint circumstellar structure. The transmission pattern of the interferometer is given by
\begin{equation}
T(\theta)=\sin^2\left(\frac{\pi\,b\,\theta}{\lambda}\right)
\end{equation}
where $b$ is the baseline of the interferometer, $\lambda$ is the wavelength of observation, and $\theta$ is the angle, along the line connecting the apertures, from the pointing center of the interferometer. Any extended object is seen in the focal plane of the interferometer as the object multiplied by the transmission pattern in equation (1) and convolved with the point-spread function of a single aperture.

Our nulling interferometer is designed for use with the converted 6.5 m MMT, based upon a prototype first demonstrated with the old Multiple Mirror Telescope (Hinz et al. 1998). The nulling instrument, called the Bracewell Infrared Nulling Cryostat (BLINC; Hinz et al. 2001), divides the pupil of the telescope into two halves and overlaps them on a 50\% transmissive beam-splitter.  The mid-infrared array camera (MIRAC; Hoffmann et al. 1998) provides a final stop for the overlapped beams defining two 2.5 m apertures separated by 4 m. At 10.3 $\mu$m this provides a constructive fringe 0.27$\arcsec$ from the star. The orientation of the interferometer can be adjusted since the MMT is an altitude-azimuth mounted telescope equipped with the ability to derotate the instrument.  The derotator stays fixed for nulling observations, with its orientation determining the interferometer baseline's position angle on the sky.  For a more detailed description see Hinz et al. (2001).

Precision nulling interferometry is designed for use with adaptive optics which corrects the atmospherically aberrated wavefront to improve the destructive interference between the beams. However, useful observations can be taken for bright objects even without atmospheric correction. When the interferometer is properly aligned and phased on an object the atmospheric phase variations between the apertures causes the image to randomly fluctuate in brightness. In effect the atmosphere acts as a phase shifter for the interferometer. By recording short exposure (0.05-0.5 s) images of the object we accumulate frames which show the star both at constructive and destructive interference. If a sequence of frames is recorded on the object, the nulled intensity measured by the instrument (instrumental null) is defined as
\begin{equation}
N_{inst}=\frac{I_{min}}{I_{max}}
\end{equation}
where I$_{min}$ and I$_{max}$ correspond to the minimum and maximum intensity of a recorded sequence. By comparing this nulled intensity to that of a calibrator star, any excess flux above that of a point source can be reliably detected down to the level of uncertainty in the measured null.  The measurements can also be thought of as measuring the visibility of the object, which is given by the relation
\begin{equation}
V_{inst}=\frac{1-N_{inst}}{1+N_{inst}}.
\end{equation}
In interferometry the true source visibility, given an instrumental visibility ($V_i$) and a calibration point source visibility ($V_c$), is defined as $V_s=\frac{V_i}{V_c}$. From this it can be shown that the true nulled flux, or source null, is given approximately as
\begin{equation}
N_s\simeq N_i - N_c.
\end{equation}

\section{Circumstellar Disk Modelling}

 Models which have been able to successfully address the SED of T Tauri and Herbig Ae stars are dominated by reprocessing of the stellar radiation via a flared disk. This creates a surface layer which is superheated by the stellar irradiation (Kenyon and Hartmann 1987; Chiang and Goldreich 1997; D'Alessio et al. 1998). For the purpose of estimating the size of mid-IR emission in such a disk we use the model of Chiang and Goldreich (1997, hereafter CG). 

\begin{figure}
\epsfxsize=\linewidth
\epsfbox{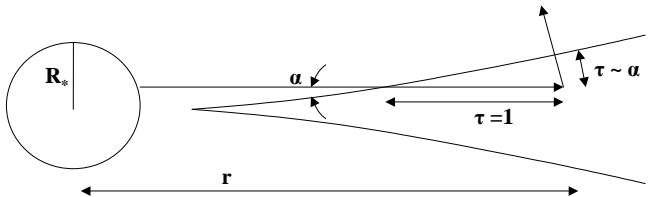}
\caption{Sketch of circumstellar disk model (Chiang and Goldreich 1997). The star irradiates the disk at grazing incidence penetrating to an optical depth of unity and creating a layer of dust which is heated above that of the disk interior.  The optical depth of the heated surface is proportional to the incidence angle of the starlight. }
\end{figure}

The spatial distribution of the IR emission due to the heated atmosphere of the disk is determined by the temperature distribution as well as the optical depth of the atmosphere.  CG derive a temperature distribution for the dust atmosphere of
\begin{equation}
T_{atm}(r)\simeq T_{\ast}\left(\frac{R_{\ast}}{2r}\right)^{2/5}
\end{equation}
where T$_\ast$ and R$_{\ast}$ are the effective temperature and radius of the central star.  If the stellar radiation strikes the disk surface at an angle, $\alpha$, as shown in Figure 1, then, from geometrical considerations, it can be seen that the optical depth of the superheated layer is proportional to $\alpha$, since the stellar radiation will penetrate to unity optical depth at grazing incidence. For a disk in hydrostatic equlibrium the grazing angle is given by CG as
\begin{equation}
\alpha=\frac{8}{7}\left(\frac{T_{\ast}}{T_c}\right)^{4/7}\left(\frac{r}{R_{\ast}}\right)^{2/7}
\end{equation}
for large distances from the star, where
$T_c=\frac{G\,M_{\ast}\mu_g}{k\,R_{\ast}}$, $\mu_g$ is the mean molecular weight of the gas, k is the Stefan-Boltzmann constant, and G is the gravitational constant.  The optical depth is of the atmosphere is then given as  $\tau=\alpha \,\epsilon_{IR}$, where $\epsilon_{IR}$ is the infrared emissivity of the material.

The interior of the disk also contributes a significant portion of the mid-IR flux. CG derive a temperature distribution for the interior of the disk of
\begin{equation}
T_{i}(r)\simeq T_{\ast}\left(\frac{\alpha}{4}\right)^{1/4}\left(\frac{R_{\ast}}{r}\right)^{1/2},
\end{equation}
which is valid for the region where the disk interior is opaque to its own radiation, true for mid-IR emission calclulations.

For an optically thin disk atmosphere and optically thick interior with temperatures as listed above, the emission contributed by a given annulus, $dr$,  of a face-on disk is 
\begin{equation}
I_{\nu}(r)=  2\pi\left(\tau(r)\,B_{\nu}(T_{atm}) + B_{\nu}(T_i)\right)\,r\,dr
\end{equation}
where B$_\nu$ is the Planck function.  For stars with parameters similar to those listed in Table 1 the 10.3 $\mu$m emission from the disk atmosphere is predicted to come from an annulus centered at $\sim$17 AU from the star.

The response of the interferometer is found by multiplying equation (8), scaled appropriately for the distance to the star, by the transmission pattern of the interferometer (equation (1)) to get
\begin{equation}
I_{null}=4\int_{0}^{r_{o}}\int_{0}^{r_{o}}I_{\nu}(r)\sin^2\left(\frac{\pi\,b \,x}{\lambda}\right)dxdy.
\end{equation}
I$_\nu$ is only nonzero for values of r $>$ 0.6 AU, corresponding to the sublimation region of silicate dust around A stars. The outer limit, r$_{o}$, is taken to be 500 AU.  The model is not sensitive to the exact value of these limits since little flux is produced at the extremes. Since the object size is on the order of the fringe spacing of the interferometer the light detected from the disk at constructive interference must also be calculated. This is given by
\begin{equation}
I_{con}=4\int_{0}^{r_{o}}\int_{0}^{r_{o}}I_{\nu}(r)\cos^2\left(\frac{\pi\,b \,x}{\lambda}\right)dxdy.
\end{equation}
The calculated expected nulls for the three Herbig Ae/Be stars are listed in Table 1.

\begin{deluxetable}{llllllr}
\tabletypesize{\small}
\tablewidth{\textwidth}
\tablecaption{Parameters and Expected Source Nulls of Herbig Ae Stars}
\tablehead{
\colhead{Star}	&\colhead{Spectral}	&\colhead{distance\tablenotemark{a}} 	&\colhead{T$_{\ast}$\tablenotemark{b}} 	&\colhead{R$_{\ast}$\tablenotemark{c}} 	&\colhead{M$_{\ast}$\tablenotemark{c}}	&\colhead{Expected}   \\
\colhead{}	&\colhead{Type}		&\colhead{(pc)}	&\colhead{(K)}		&\colhead{(R$_{\odot}$)}	&\colhead{(M$_{\odot}$)} 	&\colhead{Source Null} }

\startdata
HD150193&A2Ve		&150	&10000		&2.1		&2.6		&63\%	\\
HD163296&A0Ve		&122	&10500		&2.8		&3.3		&72\%	\\
HD179218&B9Ve		&240	&10000		&2.5		&3.4		&54\%	\\
\enddata
\tablenotetext{a}{Perryman et al. (1997).}
\tablenotetext{b}{Meeus et al. (2001).}
\tablenotetext{c}{Hillenbrand et al. (1992).}
\end{deluxetable}

\section{Observations}
Observations were taken during the commissioning run of BLINC (June 10-18, 2000) at the MMT, shortly after first light of the telescope.  Two sequences of 500 frames were recorded for each object at orthogonal orientations of the baseline (except for HD150193 which was only obtained in one orientation).  An integration time of 0.5 s was used for each frame. All objects were observed June 14, 2000 during a night of particularly good seeing.  Observations were taken using a 10.3 $\micron$, 10\% bandwidth filter. The images are sky subtracted using frames taken of adjacent sky between observations. A custom IRAF aperture photometry program is used to extract the minimum and maximum frames in a sequence to determine the instrumental null. Error in measuring the instrumental null is estimated by comparing subsequent sequences on the same object. Since the object can be detected even in the nulled image the error in estimating the null is dominated by the atmospheric variations in phase. Using this approach, measurements of the instrumental null are typically measured to a precision of several percent. 

Calibration of the observations was performed using $\alpha$ Herculi and Vega, using the same integration time as for the Herbig Ae stars. The interferometer was scanned in path-length (by moving the star across the array where an arcsecond corresponds to 38 $\micron$ of path difference) after any significant telescope movement or change in orientation which might change the relative paths of each arm and decrease the visibility.

We also observed $\alpha$ Scorpii during the same night for a different science program. The star was measured to have a source null of 19\% (visibility of 68\%) at 10.3 $\mu$m due to the extended dust outflow surrounding the star.  This measurment agrees with other mid-IR interferometric measurements such as the Infrared Spatial Interferometer (Danchi et al. 1994) and demonstrates the instrument's ability to detect extended dust structure.

\section{Discussion}
As shown in Table 2 the measured nulls for three nearby Herbig Ae stars are indistinguishable from the null measured on Vega and $\alpha$ Herculi (our point source calibrations) under the same conditions. Although the level of nulling achievable is not dramatic, due to the presence of atmospheric turbulence, the measurements clearly do not agree with the model presented of mid-IR emission for Herbig Ae stars.  

Meeus et al. (2000) present ISO spectra for a sample of Herbig Ae/Be stars including the three objects observed with nulling.  All three objects show significant silicate emission features over the 10.3 $\mu$m filter pass band used for MMT observations. The strength of emission is similar for the three stars and contributes approximately 50\% of the total flux in each case.  The relative equality of the two contributions requires that both sources of the flux, the silicate emission and continuum, be spatially unresolved, since either one, if extended, would contribute more flux to the null than is detected in the measurements. 

As pointed out by Bouwman et al. (2000), present modelling of circumstellar material to fit an object's SED is degenerate with respect to the size distribution of the grains as well as their geometric distribution. For example, larger grains closer in to the star can mimic the spectra of smaller grains further from the star, as long as the grain size distribution is consistent with the 10 $\mu$m emission feature seen by Meeus et al. (2001). A similar degeneracy exists for constraining the inner and outer boundaries of the dust region.  Thus the observations presented are best viewed in terms of breaking the current degeneracies in models of circumstellar material using the SED alone, by providing spatial constraints for the mid-IR dust emission. 

	One might think that the observations can be explained by nearly edge-on disks which shield our line of sight to the flared disk atmospheres. However the silicate emission seen by ISO suggests we are seeing optically thin irradiated dust, suggestive of a surface layer or envelope. In addition inclined models by Chiang and Goldreich (1999) predict a significantly different SED for inclinations greater than 60$^\circ$. For inclinations less than this, even the minor axis of the projected disk would be detectable with our interferometer.

	Independent of interpretation, however, the observations constrain the spatial extent of circumstellar disks around Herbig Ae stars. To fit the observations the convolution of a model flux with the transmission pattern of the interferometer must be less than the listed values in Table 2. Roughly, this corresponds to 90\% of the 10 $\micron$ flux of the disk coming from within 0.13$\arcsec$ diameter or 20 AU at a distance of 150 pc. Observations of a larger sample of Herbig Ae stars are needed to verify this result. 

\clearpage

\begin{deluxetable}{llrr}
\tabletypesize{\small}
\tablewidth{\linewidth}
\tablecaption{Results of Nulling Observations}
\tablehead{
\colhead{Star}	&\colhead{Instrumental} &\colhead{Source\tablenotemark{a}}  	&\colhead{Position}\\ 
\colhead{}	&\colhead{Null}		&\colhead{Null}				&\colhead{Angle} }
\startdata
HD150193	&13$\pm$5\%	&0$\pm5$\%	&97$^\circ$ \\
HD163296	&12$\pm$7\%	&-1$\pm$7\%	&94$^\circ$\\
HD163296	&16$\pm$2\%	&3$\pm$3\% 	&10$^\circ$\\
HD179218	&16$\pm$2\% 	&3$\pm$3\% 	&162$^\circ$\\
HD179218	&14$\pm$2\% 	&1$\pm$3\%	&87$^\circ$\\
\enddata
\tablenotetext{a}{Calibrator stars were $\alpha$ Her and Vega with measured nulls of 13$\pm$2\% and 13$\pm$3\% respectively using the same integration time as the science objects.}
\end{deluxetable}

Of the three stars, the most detailed information about the surrounding excess is known for HD163296.  Mannings and Sargeant (1997) observed it at 2.7 mm and found an elongated dust disk, at an inclination of 58$^\circ$ approximately 110 AU in diameter with a position angle of 126$^\circ$.  Grady et al. (2000) confirm this structure in the visible with HST, where a disk illuminated by scattered light is seen with an inclination of 60$^{\circ}$ and a position angle of 140$^\circ$.  The position angles corresponding to the two nulling observations for this star are 10$^\circ$ and 94$^\circ$. Although neither of these lies along the major axis of the disk, the two observations would still be expected to resolve the disk at the intermediate angles of observation if the mid-IR emission followed the spatial structure of the CG model. At an inclination of 60$^{\circ}$, if the inclination does not cause obscuring of the emitting region, the expected source null would still be 60\% if the disk major axis were oriented perpendicular to the baseline direction of the interferometer.

\section{Conclusions}
 The observations of three Herbig Ae stars using nulling interferometry constrains disks around these objects to be less than approximately 20 AU diameter in the mid-IR. Although more objects will need to be observed to confirm this conclusion, these first observations, if indicative of the sample, provide a mid-IR spatial constraint for models of proto-planetary disks around Herbig Ae stars.

These observations demonstrate that nulling interferometry is a useful technique for probing pre-main sequence circumstellar environents in the thermal IR. The constraints placed on the spatial extent of the disk, even though they were achieved using a single aperture telescope, are more stringent than is possible by using standard imaging. Typical widths of images of the models presented in \S 3 would result in an increase of roughly 20\% in the object's FWHM, compared to a point source. While nominally measurable, effects such as chopper jitter, atmospheric variations, and variable optical aberrations could all work to mimic such a small increase in image width.  In contrast the interferometer converts the observation to a relative photometric measurement, which is straightforward to measure and calibrate.

\acknowledgements
We would like to thank Marc Kuchner for suggesting the attractiveness of using nulling to probe Herbig Ae/Be stars.  PH acknowledges support from the Michelson Graduate Fellowship Program. The BLINC interferometer was developed under a NASA/JPL grant for TPF. MIRAC is supported by grant AST9618850 from the National Science Foundation.


\begin{thebibliography}{}
\bibitem[Bouwman, de Koter, van den Ancker, \& 
Waters(2000)]{2000tesa.conf...63B} Bouwman, J., de Koter, A., van den 
Ancker, M.~E., \& Waters, L.~B.~F.~M.\ 2000, ASP Conf.~Ser.~196: Thermal 
Emission Spectroscopy and Analysis of Dust, Disks, and Regoliths, 63 
\bibitem[Chiang \& Goldreich(1997)]{1997ApJ...490..368C} Chiang, E.~I.~\& 
Goldreich, P.\ 1997, \apj, 490, 368 
\bibitem[Chiang \& Goldreich(1999)]{1999ApJ...519..279C} Chiang, E.~I.~\& 
Goldreich, P.\ 1999, \apj, 519, 279
\bibitem[D'Alessio, Canto, Calvet, \& Lizano(1998)]{1998ApJ...500..411D} 
D'Alessio, P., Canto, J., Calvet, N., \& Lizano, S.\ 1998, \apj, 500, 411  
\bibitem[Danchi et al.(1994)]{1994AJ....107.1469D} Danchi, W.~C., Bester, 
M., Degiacomi, C.~G., Greenhill, L.~J., \& Townes, C.~H.\ 1994, \aj, 107, 
1469 
\bibitem[]{D} Dullemond, C.P., Dominik, C., \& Natta, A. 2001 astro-ph/0106470
\bibitem[Grady et al.(2000)]{2000ApJ...544..895G} Grady, C.~A.~et al.\ 
2000, \apj, 544, 895 
\bibitem[Hillenbrand, Strom, Vrba, \& Keene(1992)]{1992ApJ...397..613H} 
Hillenbrand, L.~A., Strom, S.~E., Vrba, F.~J., \& Keene, J.\ 1992, \apj, 
397, 613 
\bibitem[Hinz et al.(2001)]{} Hinz, P.~M., Angel, 
J.~R.~P., Woolf, N.~J., Hoffmann, W.~F., \& McCarthy, D.~W.\ 2001, in preparation 
\bibitem[Hinz et al.(1998)]{1998Natur.395..251H} Hinz, P.~M., Angel, 
J.~R.~P., Hoffmann, W.~F., McCarthy, D.~W., McGuire, P.~C., Cheselka, M., 
Hora, J.~L., \& Woolf, N.~J.\ 1998, \nat, 395, 251 
\bibitem[Hoffmann et al.(1998)]{1998SPIE.3354..647H} Hoffmann, W.~F., Hora, 
J.~L., Fazio, G.~G., Deutsch, L.~K., \& Dayal, A.\ 1998, \procspie, 3354, 
647 
\bibitem[Kenyon \& Hartmann(1987)]{1987ApJ...323..714K} Kenyon, S.~J.~\& 
Hartmann, L.\ 1987, \apj, 323, 714 
\bibitem[Mannings \& Sargent(1997)]{1997ApJ...490..792M} Mannings, V.~\& 
Sargent, A.~I.\ 1997, \apj, 490, 792 
\bibitem[Meeus et al.(2001)]{2001A&A...365..476M} Meeus, G., Waters, 
L.~B.~F.~M., Bouwman, J., van den Ancker, M.~E., Waelkens, C., \& Malfait, 
K.\ 2001, \aap, 365, 476 
\bibitem[Natta et al.(2001)]{2001A&A...371..186N} Natta, A., Prusti, T., 
Neri, R., Wooden, D., Grinin, V.~P., \& Mannings, V.\ 2001, \aap, 371, 186 
\bibitem[Natta, Grinin, \& Mannings(2000)]{2000prpl.conf..559N} Natta, A., 
Grinin, V., \& Mannings, V.\ 2000, Protostars and Planets IV, 559 
\bibitem[Perryman et al.(1997)]{1997A&A...323L..49P} Perryman, M.~A.~C.~et 
al.\ 1997, \aap, 323, L49 


\end{thebibliography}
\end{document}